\documentclass[letter,10pt]{article}
\usepackage{graphicx}
\begin{document}

\begin{flushleft}
{\sf {\Large Quantum secret sharing based on modulated high-dimensional time-bin entanglement}} 
\end{flushleft}

\begin{flushleft}
Hiroki Takesue$^{1,3}$ and  Kyo Inoue$^{1,2,3}$\\
$^1$NTT Basic Research Laboratories, NTT Corporation\\
3-1 Morinosato Wakamiya, Atsugi, Kanagawa, 243-0198, Japan\\
$^2$Department of Electrical, Electronics and Information Engineering, Osaka University\\
2-1 Yamadaoka, Suita, Osaka, 565-0871, Japan\\ 
$^3$CREST, Japan Science and Technology Agency\\
4-1-8 Honcho, Kawaguchi, Saitama, 332-0012, Japan\\
\today
\end{flushleft}

\begin{flushleft}
Abstract
\end{flushleft}
We propose a new scheme for quantum secret sharing (QSS) that uses a modulated high-dimensional time-bin entanglement. 
By modulating the relative phase randomly by $\{0,\pi\}$, a sender with the entanglement source can randomly change the sign of the correlation of the measurement outcomes obtained by two distant recipients. 
The two recipients must cooperate if they are to obtain the sign of the correlation, which is used as a secret key. We show that our scheme is secure against intercept-and-resend (I-R) and beam splitting attacks by an outside eavesdropper thanks to the non-orthogonality of high-dimensional time-bin entangled states.
We also show that a cheating attempt based on an I-R attack by one of the recipients can be detected by changing the dimension of the time bin entanglement randomly and inserting two ``vacant" slots between the packets. 
Then, cheating attempts can be detected by monitoring the count rate in the vacant slots. 
The proposed scheme has better experimental feasibility than previously proposed entanglement-based QSS schemes.

\section{Introduction}

Many quantum information systems including quantum cryptography and quantum computer have been intensively studied in recent years \cite{pqi}. 
Of these, quantum secret sharing (QSS) has been attracting attention \cite{hillery,karlsson,tittel,bagh,xiao,singh,zhang,schmid}. 
The basic idea of secret sharing is that a secret key transmitted by a sender is shared between two or more recipients in such a way that the key can be reconstructed only if all recipients collaborate. The purpose of QSS is to provide this function with absolute security using quantum mechanics : a secret key from a sender is transmitted over a quantum channel to two (or more) recipients, and the key is used to encrypt communication between the sender and the recipients in a classical channel that cannot be modified but may be overheard by an eavesdropper.

The first QSS scheme proposed by Hillery et al. used a three-particle entangled Greenberger-Horne-Zeilinger (GHZ) state \cite{hillery}. Although this scheme elegantly showed the essence of QSS, it is hard to realize experimentally because of the inefficiency as regards the generation of a three-particle entangled state \cite{bouwmeester}. 
Several variations and theoretical expansions of QSS have been reported since the publication of this pioneering work \cite{karlsson,tittel,bagh,xiao,singh,zhang,schmid}.
Among them, schemes based on two-particle entangled states seem to have good experimental feasibility with optical setups \cite{karlsson,tittel}. 
These schemes use four non-orthogonal Bell states and two measurement bases that are non-orthogonal to each other to prevent an eavesdropper from obtaining the key without inducing errors. Therefore, the experimental configurations are complex and difficult to implement. 
Recently, simpler schemes have been proposed based on sequential communication of a single qubit \cite{zhang,schmid}. 
In a sense, these methods insert some users who undertake unitary transformation into a transmission line of quantum key distribution (QKD) systems using single photons. Therefore, the secure key distribution distances of these methods are expected to be similar to those of QKD systems using single photons. 
In an analogy with the relationship between the secure key distribution distance of a single-photon-based QKD and that of an entanglement-based QKD, the secure key distribution distance of these QSS schemes will be smaller than those of entanglement-based QSS schemes. 

In this paper, we propose a new QSS scheme based on a two-photon entangled state. Our approach employs high-dimensional time-bin entanglement \cite{ried,stucki}, which is an expansion of time-bin entanglement with two time slots \cite{brendel}. 
We apply a differential phase modulation to high-dimensional time-bin entanglement.
We show that our scheme is secure against an intercept-and-resend (I-R) attack or a beam splitting (BS) attack, 
because an eavesdropper cannot reconstruct the whole wavefunction of the modulated high-dimensional time-bin entanglement by such attacks. In other words, our scheme utilizes the non-orthogonality of modulated high-dimensional entangled states to ensure that an eavesdropper cannot determine the state with a single measurement, instead of using four non-orthogonal Bell states as in \cite{karlsson,tittel}.
In addition, the dimensions of the time-bin entangled states are randomly changed packet by packet, and two vacant time slots are inserted between packets. 
As a result, cheating attempts based on an I-R attack by one of the recipients can be detected by monitoring the count rate of the vacant slots. 
Note that our security analysis is based on specific attacks, and a full security analysis to prove unconditional security is beyond the scope of this paper. 
The source of a high-dimensional entangled state is easier to construct than the sources of previous entanglement-based QSS schemes. In addition, the recipients of our scheme do not have to select one from two non-orthogonal measurement bases. These characteristics make the configuration very simple and the presented scheme experimentally feasible. 

The structure of this paper is as follows. An overview of the proposed scheme is provided in Sec. 2. In sections 3 and 4, we discuss security against eavesdropping by an outsider and cheating attempts by one of the recipients, respectively.  
In section 5 we discuss the obtained results, and describe a possible modification of the proposed protocol. We conclude the paper in Sec. 6.

\section{Proposed scheme}

\begin{figure}[thb]

\centerline{\includegraphics[width=\linewidth]{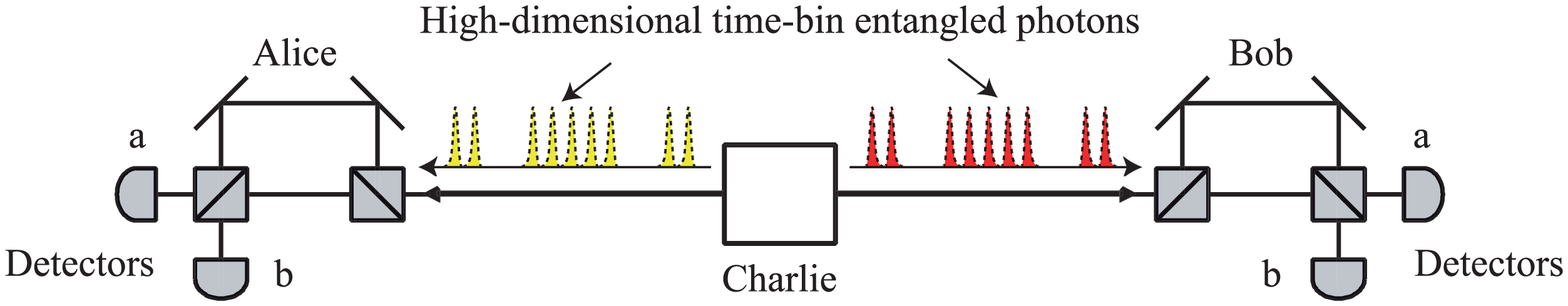}}

\caption{Schematic of the proposed QSS system.}
\label{setup}

\end{figure}

%%%%%%%%%%%%%%%%high-dimensional%%%%%%%%%%%%

Figure \ref{setup} shows the configuration of our proposed scheme, which is based on a simple Franson interferometer setup \cite{franson}. 
Charlie generates packets of high-dimensional time-bin entangled states spanned by $N$ time slots, each of which is followed by two vacant time slots. The state of a packet (with the following vacant slots) is expressed as
\begin{equation}
|\Phi\rangle = \frac{1}{\sqrt{N}} \sum_{k=1}^N e^{i \phi_k} |k\rangle_s |k\rangle_i  + 0 \cdot |N+1\rangle_s |N+1\rangle_i + 0 \cdot |N+2\rangle_s |N+2\rangle_i, \label{stime}
\end{equation}
where the expression $|k\rangle_x$ represents a state in which there is a photon in the $k$th time slot in a mode $x$, signal ($s$) or idler ($i$). 
$\phi_k$ is the phase at the $k$th time slot and is modulated randomly by $\{0,\pi\}$. 
The dimension $N$ is randomly changed packet by packet. 
The signal and idler photons are separated and sent to Alice and Bob, respectively. Alice and Bob put the photons into 1-bit delayed interferometers whose two outputs are connected to photon counters. 
A state $|k\rangle_s$ is converted as follows by a 1-bit delayed interferometer. 
\begin{equation}
|k\rangle_x \to \frac{1}{2} \left(|k,a\rangle_x - |k,b\rangle_x + |k+1,a\rangle_x + |k+1,b \rangle_x \right) \label{henkan}
\end{equation}
In the expression $|k,y\rangle_x$, $k$ shows the time slot where there is a photon, $y$ is the output port ($a$ or $b$) of the delayed interferometer, and $x$ denotes signal ($s$) or idler ($s$). 
By plugging Eq. (\ref{henkan}) into Eq. (\ref{stime}), we can obtain the state at the output of the interferometers, which is shown by 
\begin{eqnarray}
|\Phi \rangle &\to& \frac{1}{4 \sqrt{N}} \left[e^{i\phi_1}|1,a\rangle_s |1, a\rangle_i - e^{i\phi_1}|1,a\rangle_s |1,b\rangle_i - e^{i\phi_1}|1,b\rangle_s |1, a\rangle_i + e^{i\phi_1}|1,b\rangle_s |1,b\rangle_i \right.  \nonumber \\
& & + \sum_{k=2}^N \left\{ (e^{i\phi_{k-1}}+e^{i\phi_k})|k,a\rangle_s |k, a\rangle_i + (e^{i\phi_{k-1}}-e^{i\phi_k})|k,a\rangle_s |k,b\rangle_i \right. \nonumber \\
& & \left. +(e^{i\phi_{k-1}}-e^{i\phi_k})|k,b\rangle_s |k, a\rangle_i + (e^{i\phi_{k-1}}+e^{i\phi_k})|k,b\rangle_s |k,b\rangle_i \right\} \nonumber \\
& & + e^{i \phi_{N}} |N+1,a\rangle_s |N+1, a\rangle_i + e^{i \phi_{N}}|N+1,a\rangle_s |N+1,b\rangle_i \nonumber \\
& & \left. + e^{i \phi_{N}}|N+1,b\rangle_s |N+1,a\rangle_i + e^{i \phi_{N}}|N+1,b\rangle_s |N+1,b\rangle_i+ \cdots
\right], \label{1bitN}
\end{eqnarray}
where only terms that contribute to coincidence are shown.  
Thus, quantum interference is observed in the time slots from 2 to $N$, which we call ``signal slots" hereafter. 
When $\phi_{k} = \phi_{k-1}$, Alice and Bob's outcomes observed in the $k$th signal slot are positively correlated.  Anti-correlation is observed when $\phi_{k}=\phi_{k-1}\pm \pi$. Thus, we can change the sign of the correlation for each time slot by modulating differential phase $\Delta \phi_k = \phi_k - \phi_{k-1}$ by $\{0,\pi\}$ for each $k$. 
By contrast, the result of the coincidence in the 1st or $(N+1)$th slots is completely random. We call these slots ``error slots". 
Alice and Bob do not observe any count in the $(N+2)$th slot. As described in Sec. 4, the $(N+2)$th slot is used to detect cheating attempts by participants, so we call this slot a ``detection slot".
Our QSS depends on the fact that the sign of the correlation in a signal slot is determined only by summarizing the outcomes of both Alice and Bob.

%%%%%%%%%%%%%%%%%%%procedure%%%%%%%%%%%%%%%%%%%%
Using this characteristic, we can realize a QSS with the following procedure. For simplicity, we do not consider the vacant slots. The procedure for detecting cheating using the count rate in the detection slots is described in Sec. 4. 

\begin{itemize}
\item[1.] Charlie generates entangled photon pairs whose state is given by Eq. (\ref{stime}), with $\phi_k$ modulated by $\{0,\pi\}$. He separates signal and idler photons and sends them to Alice and Bob.
\item[2.] Alice and Bob input the received photons into their interferometers, and detect the photons with photon counters connected to two output ports of the interferometers. Alice and Bob record the time instances in which they observed clicks, and which detectors clicked for each time instance. 
\item[3.] Alice and Bob inform Charlie of the time instances in which they observed clicks via classical communication. They do not disclose the detectors that clicked. 
\item[4.] Charlie makes a key sequence using his modulation data in signal slots. Charlie encodes a message using this key and sends it to Alice and Bob. 
\item[5.] Charlie discloses the positions of the signal, error and detection slots to Alice and Bob. 
\item[6.] Alice and Bob discover the key only when they investigate the sign of the correlation in the signal slots by combining their information. 
\end{itemize}

%%%%%%%%%%%%%%%implementation%%%%%%%%%%%%%%%%%%%%%%%

\begin{figure}[thb]

\centerline{\includegraphics[width=\linewidth]{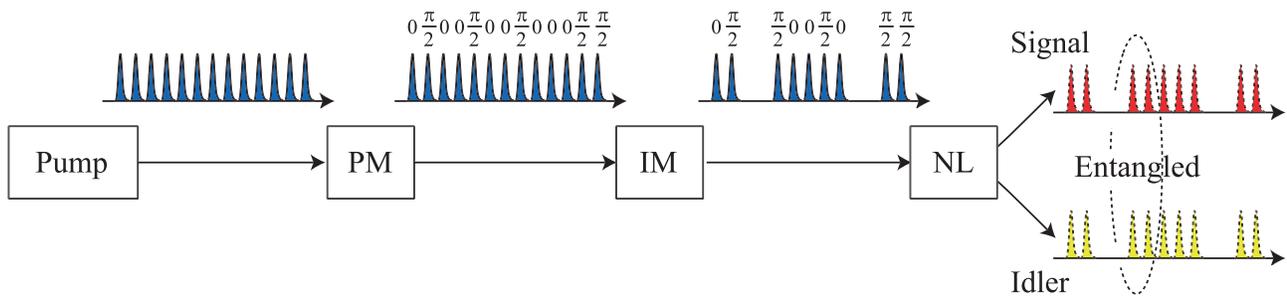}}

\caption{High-dimensional time-bin entanglement source with relative-phase modulation. PM: phase modulator, IM: intensity modulator, NL: nonlinear medium for spontaneous four-wave mixing. }
\label{source}

\end{figure}

We can generate the high-dimensional entangled state shown by Eq. (\ref{stime}) by using spontaneous parametric down-conversion (SPDC) \cite{kwiat1,kwiat2} or spontaneous four-wave mixing (SFWM) \cite{fio,takesue1,li,takesue2} with a modulated pump. An example of such an entanglement source is shown in Fig. \ref{source}, in which SFWM is used to generate time-bin entangled photon pairs. 
A coherent pulse train from a pump pulse source is launched into a phase modulator, which modulates the phase of each pulse by $\{0,\pi/2\}$. Then on-off modulation is applied to the pulse train to insert two-sequential vacant slots randomly. Thus we can generate $N$-sequential pump pulses with $\{0,\pi/2\}$ differential phase modulation, followed by two vacant slots. These pump pulses are input into a nonlinear medium, in which entangled photon pairs are generated through the SFWM process with a degenerated pump. The relationship between the phases of pump $\phi_p$, signal $\phi_s$ and idler $\phi_i$ is given by
\begin{equation}
2 \phi_p = \phi_s + \phi_i.
\end{equation}
With the above condition satisfied, and if we set the pump power relatively small so that the average number of photon pairs per slot becomes $<1$, the obtained state can be approximated as a high-dimensional entangled photon pair whose relative phase is modulated by $\{0,\pi\}$ as in Eq. (\ref{stime}). 

With these types of sources, the distribution of the number of photon pairs per packet becomes Poissonian. This implies that it is possible for two or more photons to be available in a packet, which opens the possibility for an eavesdropping strategy such as a BS attack. However, such an attack cannot be effective as long as the average number of photon pairs per pulse is sufficiently small, as we show in the next section. 

%%%%%%%%%%%%%%%%%
The previous QSS schemes based on entanglement use both the sign of correlation and the outcome of a phase-difference measurement of each photon (i.e. which detector clicked at the recipients' sites) \cite{hillery,karlsson}. In contrast, our scheme uses only the sign of correlation, which makes our approach much simpler than previous schemes. 
Moreover, recipients do not require equipment for selecting two non-orthogonal measurement bases as in \cite{hillery,karlsson}. This is not a trivial issue in terms of implementation, because such equipment usually reduces the key distribution distance. For example, the use of active components such as a phase modulator for basis selection induces additional loss, which reduces the loss budget for transmission. 
Despite its simplicity, our scheme is secure against eavesdropping from outside and cheating attempts by one of the recipients, as described in the next two sections.

\section{Eavesdropping by outsider}

In this section we analyze the security of our QSS scheme against possible eavesdropping from outside the party. 
An eavesdropper (Eve) can undertake attacks that are similar to those against QKD systems, such as an I-R attack or a BS attack. In the following, we consider these two types of attack.

\subsection{I-R attack}

Eve captures both photons from Charlie and undertakes coincidence measurements using similar interferometers to those of Alice and Bob to obtain the relative phase $\Delta \phi_k=\phi_k - \phi_{k-1}$. However, she can obtain only partial information about the relative phases of high-dimensional time-bin entangled states, because the average number of photon pairs per pulse is smaller than one. 

Then, at the time instances where Eve obtained $\Delta \phi_k$, she prepares substitute photons in which the relative phase information that she obtained is encoded and sends them to Alice and Bob through lossless lines.
Eve has two choices of substitute photons: entangled photon pairs or pairs of single photons with classical correlation. 

First, we consider the case in which Eve prepares entangled photon pairs.  
With her obtained relative phase information $\Delta \phi_k$, she generates the following entangled photon pair, splits them into signal and idler, and sends one to Alice and the other to Bob. 
\begin{equation}
|\Phi_e\rangle = \frac{1}{\sqrt{2}} (|k-1\rangle_s |k-1\rangle_i + e^{i\Delta \phi_k} |k\rangle_s |k\rangle_i)
\end{equation}
When the above photon pair passes the interferometers of Alice and Bob, whose function is given by Eq. (\ref{henkan}), the state is converted into
\begin{eqnarray}
|\Phi_e\rangle &\to& \frac{1}{4 \sqrt{2}} \left\{ |k-1,a\rangle_s |k-1,a\rangle_i  - |k-1,a\rangle_s |k-1,b\rangle_i \right. \nonumber \\
& & - |k-1,b\rangle_s |k-1,a\rangle_i  +|k-1,b\rangle_s |k-1,b\rangle_i  \nonumber \\
& & +(1+e^{i\Delta \phi_k}) |k,a\rangle_s |k, a\rangle_i + (1-e^{i\Delta \phi_k})|k,a\rangle_s |k,b\rangle_i \nonumber \\
& & + (1-e^{i\Delta \phi_k})|k,b\rangle_s |k, a\rangle_i + (1+e^{i\Delta \phi_k})|k,b\rangle_s |k,b\rangle_i \nonumber \\
& & e^{i\Delta \phi_k} |k+1,a\rangle_s |k+1,a\rangle_i  + e^{i\Delta \phi_k} |k+1,a\rangle_s |k+1,b\rangle_i \nonumber \\
& & \left. + e^{i\Delta \phi_k} |k+1,b\rangle_s |k+1,a\rangle_i  +e^{i\Delta \phi_k} |k+1,b\rangle_s |k+1,b\rangle_i + \cdots \right\}. \label{eve}
\end{eqnarray}  
where only terms that contribute to coincidences are shown. 
The above equation indicates that Alice and Bob possibly observe coincidences in either of the $k-1$, $k$, and $(k+1)$th slots. If they observe coincidences in the $k$th time slot, their measurement result correlates with Charlie's phase modulation data, which means that the eavesdropping is successful. 
However, when they observe a coincidence in the $(k\pm 1)$th slots, their outcomes are uncorrelated, so error occurs with a 50\% probability. From the probability amplitude of Eq. (\ref{eve}), an error occurs with 1/4 probability for a photon pair resent by Eve. 

Next, we consider the case in which Eve uses classical correlation. She prepares two single photons. Each photon is made into a 2-slot time-bin qubit whose relative phase is modulated based on the relative phase $\Delta \phi_k$ that she obtained in her measurement. The joint state of the two photons is expressed as
\begin{equation}
|\Phi_c\rangle = \frac{1}{2} (|k-1\rangle_A + e^{i \phi_{Ak}} |k\rangle_A) (|k-1\rangle_B + e^{i \phi_{Bk}} |k\rangle_B).
\end{equation}
where $\phi_{Ak}=\phi_{Bk}$ when $\Delta \phi_k = 0$ and $\phi_{Ak} = \phi_{Bk} + \pi$ when $\Delta \phi_k = \pi$. With this condition, Eve changes $\phi_{Ak}$ and $\phi_{Bk}$ randomly by $\{0,\pi\}$.  
Eve inputs these two photons into Alice and Bob's interferometers. After these photons pass through the interferometers, the whole state changes to 
\begin{eqnarray}
|\Phi_c\rangle &\to& \frac{1}{8} \left\{ |k-1,a\rangle_A |k-1,a\rangle_B  - |k-1,a\rangle_A |k-1,b\rangle_B \right. \nonumber \\
& & - |k-1,b\rangle_A |k-1,a\rangle_B  +|k-1,b\rangle_A |k-1,b\rangle_B  \nonumber \\
& & (1 +e^{i \phi_{Ak}})(1 +e^{i \phi_{Bk}}) |k,a\rangle_A |k, a\rangle_B + (1 +e^{i \phi_{Ak}})(1 -e^{i \phi_{Bk}})|k,a\rangle_A |k,b\rangle_B \nonumber \\
& & + (1 -e^{i \phi_{Ak}})(1 +e^{i \phi_{Bk}})|k,b\rangle_A |k, a\rangle_B + (1 -e^{i \phi_{Ak}})(1 -e^{i \phi_{Bk}})|k,b\rangle_A |k,b\rangle_B \nonumber \\
& & e^{i(\phi_{Ak}+\phi_{Bk})} |k+1,a\rangle_A |k+1,a\rangle_B  + e^{i(\phi_{Ak}+\phi_{Bk})} |k+1,a\rangle_A |k+1,b\rangle_B \nonumber \\
& & \left. + e^{i(\phi_{Ak}+\phi_{Bk})} |k+1,b\rangle_A |k+1,a\rangle_B  +e^{i(\phi_{Ak}+\phi_{Bk})} |k+1,b\rangle_A |k+1,b\rangle_B + \cdots \right\}, \label{eve2} 
\end{eqnarray}
where non-coincident terms are not shown for simplicity. 
When Alice and Bob observe coincidences in the $k$th time slot, the result is correlated to Charlie's modulation data as in the previous case, which means the eavesdropping is successful. However, the coincidences obtained in the $(k \pm 1)$th slot induce errors with 50\% probability. Using the probability amplitude of Eq. (\ref{eve2}), the error probability is calculated to be 1/6. This means that using two single photons with classical correlation is the better strategy for Eve. 
In either case, the I-R attack by Eve can be detected by monitoring errors with some test bits.

\subsection{BS attack}

As stated in the previous section, if we use an entanglement source based on a parametric process pumped by a coherent source, the distribution of the number of photon pairs becomes Poissonian. This makes it possible for Eve to obtain information on the keys using a BS attack, without inducing errors. Here we show that a BS attack is ineffective against our scheme with a Poissonian photon-pair source, if the average number of photon pairs per slot is fewer than 1.

We assume that the fiber transmittance between Charlie and Alice and that between Charlie and Bob are both $\alpha$, and the average number of photon pairs per slot is $\mu$. 
Eve replaces the fibers with her lossless lines. She then splits Charlie's signal and idler photon output into two paths with a beam splitter. Eve sends one beam with an average photon number of $\mu \alpha$ per pulse to both Alice and Bob through her lossless fiber so that they do not notice any eavesdropping from changes in the count rates. 
Eve keeps the other beams of signal and idler photons with an average photon number of $\mu(1-\alpha)$ in her quantum memory. Once Alice and Bob have disclosed the time instances in which they observed clicks, Eve puts her photons into interferometers that are identical to Alice and Bob's, and observes the coincidences, hoping that she obtains coincidences at the same time instances in which Alice and Bob observed coincidences. The probability of Eve obtaining a coincidence at a desired time instance is $\frac{1}{2} \mu(1-\alpha)^2$ \cite{memo}, which is close to $\frac{1}{2} \mu$ when $\alpha$ is small. Therefore, in terms of the total sifted keys of $n_{sif}$ bits, Eve has full information on at most $\frac{1}{2} \mu n_{sif}$ bits. This suggests that Eve can obtain only a fraction of the information if $\mu<1$, and so Alice, Bob and Charlie can ensure that Eve's mutual information is negligible by privacy amplification \cite{lutken}.

Thus, our QSS scheme is robust against attacks by an outside eavesdropper, even with a Poissonian photon-pair source. From the above arguments, it is clear that the robustness comes from the fact that Eve cannot reconstruct the whole wavefunction of a relative-phase modulated high-dimensional time-bin entanglement by an I-R attack or a BS attack.

\section{Dishonest Bob}

\subsection{Bob's strategy}
In a QSS system, the sender of a secret key is a fair person as regards all recipients and does not have a preference for any individual, because dividing the secret key among the recipients is the purpose of the sender. 
Although an outside takeover of Charlie is a threat, this attack can be prevented by introducing an initial authentification procedure among the participants as in a QKD \cite{gisin}. 
However, each recipient may, if he or she has a chance, take advantage of the others and try to obtain full information on the key by himself or herself. Therefore, preventing cheating attempts by recipients is an essential function for a QSS system. In the following, we assume that Bob is dishonest, and is trying to obtain full information on the key without being noticed by Alice and Charlie using an I-R-attack based strategy. 

%%%%%%%%%%%%%%without vacant slots%%%%%%%%%%%%%%%%%%%%%%%
``Vacant slots" play a crucial role in the detection of dishonest Bob. 
First, we show that a cheating attempt by one of the recipients can constitute a very strong attack against a QSS protocol using $\{0,\pi\}$-modulated high-dimensional entanglement without vacant slots. 

Bob intercepts both photons output by Charlie, and inserts them into his interferometers for coincidence measurement. As a result, he obtains partial information about the relative phases of the high-dimensional time-bin entanglement. Then, Bob prepares a time-bin qubit using his measurement data, which is expressed by
\begin{equation}
|\psi\rangle = \frac{1}{\sqrt{2}}(|k-1\rangle + e^{i\phi_{Ak}} |k\rangle),
\end{equation}
where $\phi_{Ak} = \Delta \phi_k - \phi_{B}$ and $\phi_{B}$ is a value randomly chosen from $\{0,\pi\}$. 
Bob sends the above state to Alice through a lossless line. The state is converted to the following state by Alice's interferometer. 
\begin{eqnarray}
|\psi\rangle &\to& \frac{1}{2 \sqrt{2}}\{|k-1,a\rangle - |k-1,b\rangle +(1+e^{i\phi_{Ak}}) |k,a\rangle + (1-e^{i \phi_{Ak}}) |k,b\rangle \nonumber \\
& & + e^{i \phi_{Ak}} |k+1,a\rangle + e^{i\phi_{Ak}}|k+1,b\rangle\}
\end{eqnarray}
If Alice observes clicks in the $k$th time slot, the result correlates with Bob and Charlie's, so Bob's attack is successful. On the other hand, her result is uncorrelated if the click occurs in the $(k\pm 1)$th slots. However, unlike an I-R attack by an outsider, Bob can declare clicks only at the time instances for which he has information, and so he can avoid inducing errors. As a result, Bob can obtain full information about the key without being noticed by Alice or Charlie. Thus, dishonest Bob can determine the time instances where coincidences occur, which makes cheating by a key recipient a serious threat to this protocol, if we do not introduce the vacant slots which we describe in the next section. 

\subsection{Detection of dishonest Bob}

%%%%%%%%%%%%%%%%%%%%DB detection single coincidence%%%%%%%%%%%%%%%%%%%%%%%%%%
We consider a case in which Charlie sends states with vacant slots shown by Eq. (\ref{stime}). As we have already stated, Alice observes no count in the detection slot if there is no eavesdropping and both Alice and Bob are honest. If dishonest Bob measures both photons using two interferometers, the state observed by Bob is shown by Eq. (\ref{1bitN}). He observes coincidences in the error slot with a probability of $1/(2N)$, and the results obtained in these coincidence events have no correlation with Charlie's modulation data. However, he cannot distinguish if a coincidence is obtained in signal or error slots, so he inevitably resends a substitute photon to Alice based on his measurement results, which are possibly uncorrelated to Charlie's data. 
Here, we assume that Bob observed a coincidence in the $(N+1)$th error slot. Then, Bob believes that he observed the state shown by
\[
\frac{1}{\sqrt{2}}(|N\rangle_s |N\rangle_i + e^{i \Delta \phi_{e}} |N+1\rangle_s |N+1\rangle_i),
\]
where $\Delta \phi_e$ is the ``phase difference" observed in Bob's coincidence measurement and is actually a random value of $\{0,\pi\}$. 
Consequently, Bob generates the following state and sends it to Alice. 
\[
|\psi\rangle = \frac{1}{\sqrt{2}}(|N\rangle + e^{i \phi_{Ae}} |N+1\rangle)
\]
Here, $\phi_{Ae} = \Delta \phi_e - \phi_{Be}$ and $\phi_{Be}$ is a value randomly chosen from $\{0,\pi\}$. 
If this state passes through Alice's interferometer, the state is converted to 
\begin{eqnarray}
|\psi\rangle &\to& \frac{1}{2 \sqrt{2}}\{|N,a\rangle - |N,b\rangle +(1+e^{i \phi_{Ae}}) |N+1,a\rangle + (1-e^{i \phi_{Ae}}) |N+1,b\rangle \nonumber \\
& & + e^{i \phi_{Ae}} |N+2,a\rangle + e^{i \phi_{Ae}}|N+2,b\rangle\}.
\end{eqnarray}
This equation shows that Alice observes clicks in the detection slots with a probability of 1/4. Thus, counts in the detection slots imply cheating by the other participant, so Alice and Charlie stop the communication. 

Alice and Charlie also detect cheating by Bob if he resends a photon based on a coincidence measurement in the first slot. In this case, an erroneous count occurs in the detection slot of the previous packet.

This detection method is based on the count rate measured by a single participant, not on the coincidence, which is why we can detect cheating by a participant who can determine the time instances of coincidence events.

%%%%%%%%%%%%%%%%%%%%%%%%%%%N sequential %%%%%%%%%%%%%%%%%%%%%%%%%%%%%%%%%%%
To reduce the probability of inducing counts in the detection slots, Bob can make an attack based on sequential coincidence events. 
If he observes $n$ sequential coincidences, he resends a time-bin qubit spanned by $n+1$ slots to Alice. In this case too, there is a finite probability that Bob observes sequential coincidences that include one in the first or the $(N+1)$th error slots as the first or last coincidences in the sequence, respectively. 
Here, we assume that Bob observed $x$ sequential coincidences with the $(N+1)$th slot as the last event. 
Then, he considers that he observed the following entanglement. 
\[
\frac{1}{\sqrt{n+1}}\left(\sum_{k=1}^n e^{i \phi_{N+1-k}} |N+1-k\rangle_s |N+1-k\rangle_i + e^{i \phi_e} |N+1\rangle_s |N+1\rangle_i \right)
\]
$\phi_e$ is the ``phase" that Bob thinks he observed in the coincidence in the $(N+1)$th error slot and that is actually a random value of $\{0,\pi\}$. 
Therefore, he creates the following state and sends it to Alice. 
\begin{equation}
|\psi\rangle = \frac{1}{\sqrt{n+1}} \left( \sum_{k=1}^n e^{i \phi'_{N+1-k}} |N+1-k\rangle + e^{i \phi_e} |N+1\rangle
\right). 
\end{equation}
Here, $\phi'_{N+1-k}$ is the phase that correlates with Bob's measurement outcome in a signal slot. 
The above state is converted to the following state by Alice's interferometer. 
\begin{eqnarray}
|\psi\rangle &\to& \frac{1}{2 \sqrt{n+1}}\left[e^{i \phi'_{N+1-n}} |N+1-n,a\rangle - e^{i \phi'_{N+1-n}} |N+1-n,b\rangle \right. \nonumber \\
& & +\sum_{k=0}^{n-2} \left\{(e^{i \phi'_{N+1-n+k}}+e^{i \phi'_{N+2-n+k}}) |N+2-n+k,a\rangle \right. \nonumber \\
& & \left. + (e^{i \phi'_{N+1-n+k}}-e^{i \phi'_{N+2-n+k}}) |N+2-n+k,b\rangle \right\} \nonumber \\
& & + (e^{i\phi'_N} + e^{i\phi_{e}}) |N+1,a\rangle +(e^{i\phi'_N} - e^{i\phi_{e}}) |N+1,b\rangle \nonumber \\
& & \left. + e^{i\phi_{e}} |N+2,a\rangle + e^{i\phi_{e}}|N+2,b\rangle \right]
\end{eqnarray}
Thus, Alice observes clicks in the detection slot with a probability of $\frac{1}{2(n+1)}$, so cheating by Bob is disclosed. 
A similar analysis can be undertaken for a case where Bob observes $n$ sequential coincidences with the first slot of a packet as the first event of the sequential coincidences.

\subsection{Calculation}

If the detectors are ideal ones with no dark count, Alice and Charlie can always detect dishonest Bob by increasing the measurement time infinitely. However, in a realistic experimental situation, the ability to detect dishonest Bob is limited by the finite dark count rates of the detectors. In order to detect cheating, the count rate in the detection slot should be larger than the dark count rate in the presence of dishonest Bob. Here, we estimate the maximum key distribution distance over which dishonest Bob is detectable, assuming realistic dark count rates, and assuming that Charlie is equipped with an entanglement source whose photon-pair number distribution is Poissonian.  

We consider the state after the interferometer given by Eq. (\ref{1bitN}), and assume that the probability that a slot is a signal slot is $S$, then the probabilities that a slot is an error slot, $p_e$, or a detection slot, $p_d$ are expressed by 
\begin{eqnarray}
p_e &=& \frac{2}{3}(1-S), \\
p_d &=& \frac{1}{3}(1-S).
\end{eqnarray}

If we ignore the first and last slots of the whole session, the mark ratio (the probability that a slot is not a vacant slot) at the output of Charlie, $M$, is expressed as
\begin{equation}
M = S+\frac{p_e}{2} = \frac{1}{3}(2 S+1). \label{M}
\end{equation}
When the transmittance between Charlie and Alice/Bob is $\alpha$ and the average number of photon pairs per pulse $\mu$ is small enough to allow us to disregard accidental coincidences, the shared key rate between Alice and Bob ($=$ coincidence rate), $R_k$, is given by \cite{memo},
\begin{equation}
R_k = \frac{1}{2}\mu S \alpha^2. \label{key}
\end{equation}

%%%%%%%%\subsection{単発の同時計数によるresend}%%%%%%%%%%%%

Let us consider a case in which dishonest Bob resends a state based on a single coincidence event. 
The coincidence rate in a signal slot $R_s$ and an error slot $R_e$ are expressed as
$R_s = \frac{1}{2} \mu$ and 
$R_e = \frac{1}{4} \mu$, respectively \cite{memo}. 
The coincidence rate per slot observed by dishonest Bob is given by
\begin{equation}
R_{coin} = R_s S + R_e p_e. \label{douji1}
\end{equation}
If the following condition is satisfied, Alice can detect dishonest Bob from the reduction in the count rate.  
\begin{equation}
R_{coin} < M \mu \alpha \label{cond}
\end{equation}
Here, $\alpha$ shows the transmittance between Alice and Charlie, including the quantum efficiencies of Alice's detectors. With Eqs. (\ref{M}), (\ref{douji1}) and (\ref{cond}), the minimum transmittance for which Alice can detect dishonest Bob from the count rate decrease is expressed as
\begin{equation}
\alpha_{min} = 0.5. \label{cond+}
\end{equation}
Thus, if $\alpha$ is smaller than 0.5, it is impossible to construct a QSS system that is secure against dishonest Bob without employing the method described above. 

%Assume the total number of slots of the whole session is $N_{all}$. 
When $\alpha < \alpha_{min}$, Bob randomly chooses $M \mu \alpha$ events per slot from the coincidences he observes with a rate of $R_{coin}$, creates substitute photons, and sends them to Alice. 
The probability that Bob happens to choose a coincidence in an error slot and resends a photon, $y$, is expressed by
\begin{equation}
y=\frac{R_e p_e}{R_s S + R_e p_e} = \frac{1-S}{2 S+1}. \label{yyy}
\end{equation}
As we have already stated, when a resent photon is created based on the coincidence in an error slot, a photon is detected in a detection slot with a probability of 1/4. Therefore, the count rate in a detection slot is expressed as
\[
\frac{M \mu \alpha y}{4}.
\]
If the dark count rate of Alice's detector is $d$, the following condition must be satisfied to detect cheating. 
\begin{equation}
\frac{M \mu \alpha y}{4}>d
\end{equation}
Thus, the threshold value of the transmittance above at which dishonest Bob can be detected is expressed as. 
\begin{equation}
\alpha_{th} = \frac{12 d}{\mu(1-S)}
\end{equation}
When $\alpha$ is smaller than both $\alpha_{min}$ and $\alpha_{th}$, 
Alice cannot detect dishonest Bob. 
When we assume $d=10^{-5}$, $\mu=0.1$ and $S=0.5$, $\alpha_{th}\simeq -26$ dB gives the minimum transmittance with which detect dishonest Bob is detectable.

%%%%%%%%%%%%\subsection{$n$連の同時計数によるresend}

If Bob observes sequential coincidences and resends a state that is constructed using his measurement results, he may be able to reduce the probability of causing erroneous counts in the detection slots. 
So, we now consider a case in which Bob resends a state spanned by $n+1$ slots only when he observes $n$ sequential coincidence counts.

We assume that the total number of slots in the whole session is $N_{all}$. 
When the packet dimensions are larger than $n$ and the probability of observing a coincidence in a signal slot $R_s$ is small, $N_s$, the total number of $n$ sequential coincidences in signal slots in the whole session is approximated by 
\begin{equation}
N_{s} \simeq R_s^n S N_{all} = \frac{S \mu^n}{2^n} N_{all}. \label{rsn}
\end{equation}
Similarly, $N_e$, namely the total number of $n$ sequential coincidences, which includes $n-1$ signal slots and a error slot as the first or last event, is approximated by
\begin{equation}
N_{e} \simeq R_e R_s^{n-1} p_e N_{all}=\frac{p_e \mu^n}{2\cdot 2^n} N_{all}. \label{ren}
\end{equation}
Then, $n$-sequential coincidence rate observed by dishonest Bob, which is defined as the number of $n$-sequential coincidences normalized by the number of slots in the whole session, is expressed as
\begin{equation}
R_{coin} = \frac{N_s+N_e}{N_{all}} = R_s^n S + R_e R_s^{n-1} p_e \label{douji2}
\end{equation}
As in the single coincidence case, Bob should satisfy the condition expressed by Eq. (\ref{cond}), so as not to be detected by the reduction in the count rate at Alice's site. Using Eq. (\ref{douji2}), the minimum transmittance that violates this condition, $\alpha_{min}$ is given by
\begin{equation}
\alpha_{min} = \frac{\mu^{n-1}}{2^n} \label{loss}
\end{equation}
If the transmittance is smaller than $\alpha_{min}$, Bob can disguise Alice's count rate by using resent states generated using $n$-sequential coincidence counts.

When $\alpha < \alpha_{min}$ is satisfied, Bob randomly chooses $n$-sequential events from all the coincidence events, constructs the states using those events, and sends them to Alice. Here, he has to keep Alice's count rate at $M \mu \alpha$. 
The probability that Bob happens to choose an $n$ sequential coincidence that includes an error slot is the same as Eq. (\ref{yyy}). 
When Bob sends a state that is constructed using an $n$ sequential coincidence with an error slot, the probability that Alice detects a photon in a detection slot is given by $\frac{1}{2(n+1)}$. In order to detect dishonest Bob, the count rate at the detection slot should be larger than the dark count rate. This condition is expressed as
\begin{equation}
\frac{M \mu \alpha y}{2 (n+1)} \ge d.
\end{equation}
Using the above equation, the threshold value of the transmittance at which Alice can detect dishonest Bob is given by
\begin{equation}
\alpha_{thn} = \frac{6d (n+1)}{\mu (1-S)}. \label{thresh}
\end{equation}

\begin{figure}[thb]

\centerline{\includegraphics[width=.6\linewidth]{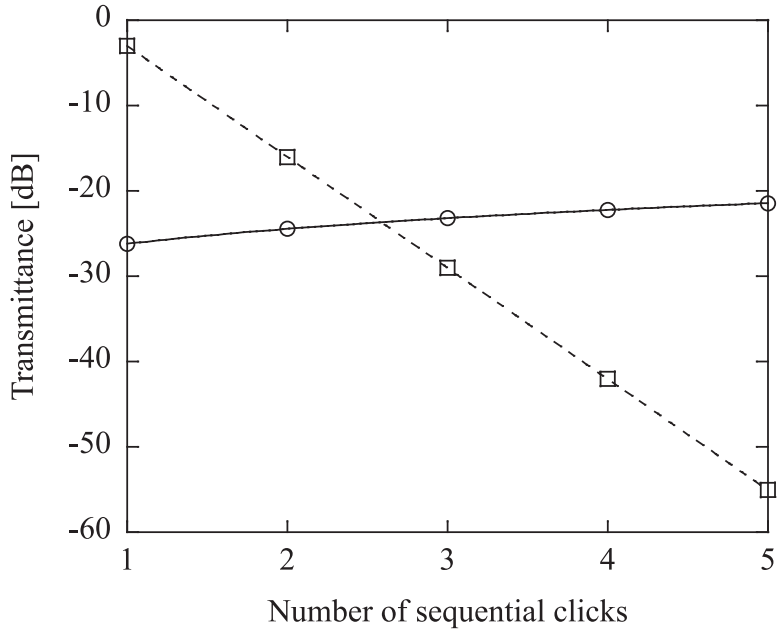}}

\caption{Threshold transmittance as a function of the number of sequential coincidences $n$. Squares show the transmittance given by Eq. (\ref{loss}), while circles show the boundary given by Eq. (\ref{thresh}). When the transmittance is smaller than both of these values, dishonest Bob can deceive Alice without being noticed. }
\label{cal}

\end{figure}

The circles in Fig. \ref{cal} show $\alpha_{thn}$ for each $n$. Here, $S$, $\mu$ and $d$ are assumed to be 0.5, 0.1 and $10^{-5}$, respectively. 
The transmittance threshold given by Eq. (\ref{loss}) is shown with squares in Fig. \ref{cal}. If the transmittance is below both curves, Bob can cheat Alice. With the above parameter values, a sequential attack with $n=2$ gives the minimum transmittance for security, $\simeq -24$ dB. When the detector quantum efficiency and out-coupling loss of the entanglement source are 10\% and 4 dB, respectively, the maximum transmission loss between Charlie and Alice/Bob is 10 dB. This means that a secure QSS system can be constructed over 100 km (50 km $\times$ 2) of optical fiber with a loss of 0.2 dB/km. 

%%%%%%%%%%%%PA etc.
This scheme uses the count rate in the detection slots to detect cheating by the key recipients. This cheat-detection method is uncommon in most of the conventional QKD or QSS systems, in which error rate monitoring is the way to detect eavesdropping and cheating. 
Therefore, a more detailed security analysis, including the derivation of the privacy amplification factor that takes account of this cheat-detection method, is very important in terms of evaluating the performance of a QSS system based on the proposed scheme.

\section{Discussion}

\subsection{Analogy with differential phase shift QKD}
 
In previously reported QSS schemes using two-photon entanglement \cite{karlsson,tittel}, entanglement sources generate four non-orthogonal Bell states and the recipients use two non-orthogonal measurement bases. Therefore, these schemes can be considered to be two-photon versions of BB84 QKD \cite{bb84}. 
In contrast, our proposed scheme can be regarded as a two-photon version of differential-phase-shift (DPS) QKD \cite{kyo02,kyo03}. 
In a DPS-QKD system, Alice randomly modulates the phase of a weak coherent pulse train by $\{0,\pi\}$ for each pulse, and sends it to Bob with an average photon number of $<1$. The state sent by Alice is expressed as
\begin{equation}
|\Phi\rangle = \frac{1}{\sqrt{N}} \sum_k^N e^{i \phi_k} |k\rangle \label{dps}
\end{equation}
where, $\phi_k =\{0,\pi\}$. 
Bob measures the phase difference of each consecutive pulse, $\Delta \phi_k = \phi_{k-1}-\phi_k$, using second-order interferometry obtained with a 1-bit delay interferometer. 
The security of the DPS-QKD is based on the fact that Eve can obtain only partial information on the relative phases in Eq. (\ref{dps}). 
On the other hand, the QSS scheme measures the relative phase of each consecutive product state shown in Eq. (\ref{stime}) using fourth-order interferometry. 
Thus, we can expand the concept of DPS-QKD to two-photon states and construct a simpler QSS system that requires only two-value phase modulation.

%%%%%%%%%modification: randomize number of spaces
\subsection{Modified scheme}

The essential point as regards detecting dishonest Bob is that the two recipients do not have the information on the time instances of the error slots. Therefore, the same function can be implemented by randomly changing the time intervals of the packets, while fixing the dimension of the packets. 
In this case, we use the same state as Eq. (\ref{stime}) with a fixed $N$, and insert random numbers of vacant time slots between packets. Then, the theory described above can be applied to this modified scheme. 
However, the scheme with randomized packet dimensions obviously utilizes time slots efficiency, so a larger key rate is expected.

\section{Conclusion}

We have proposed a new QSS scheme based on time-bin entanglement whose relative phases and dimension are modulated. A sender can change the signs of correlation of the two recipients by modulating the relative phase with $\{0,\pi\}$, which can be used as a shared key. 
This protocol is secure against an I-R attack and a BS attack from outside because of the non-orthogonality of the high-dimensional time-bin entanglement. A cheating attempt based on an I-R attack by one of the recipients can be detected by randomly changing the dimension of entanglement and inserting two vacant slots between each packet of entangled states. Because recipients cannot know the time instances of the error slots, cheating by a recipient induces an erroneous count in a detection slot, by which the other recipient can detect the existence of cheating. 
We analyzed the security of the proposed protocol based on specific attacks, and so an unconditional security analysis is an important future consideration.
This scheme does not require a three-particle entangled state or basis selection mechanism at a recipients' site, and thus has better experimental feasibility. 

This work was supported in part by National Institute of Information and Communications Technology (NICT) of Japan.


\begin{thebibliography}{99}
\bibitem{memo} Equation (\ref{1bitN}) implies that, if transmittance between Charlie and Alice/Bob is 1, the coincidence rates in a signal and an error slot are obtained as $\frac{1}{2N}$ and $\frac{1}{4N}$, respectively. 
When a Poissonian photon-pair source with the average number of photon pairs per pulse $\mu$ is used, the average number of photon pairs in a packet, whose state is approximated by Eq. (\ref{stime}), is $\mu N$. Therefore, the probabilities of a coincidence in a signal and an error slot are given by $\frac{\mu}{2}$ and $\frac{\mu}{4}$, respectively.
\end{thebibliography}
\end{document}